\documentclass[
reprint,
 aps,pr1
]{revtex4-2}
\usepackage{graphicx,color}% Include figure files
\usepackage{dcolumn}% Align table columns on decimal point
\usepackage{bm}% bold math
\usepackage{hyperref}% add hypertext capabilities
\usepackage[mathlines]{lineno}% Enable numbering of text and display math
\usepackage{physics}
\usepackage{multirow}
\usepackage{ragged2e}
\usepackage{amsmath}
\usepackage{amsfonts}
\usepackage{comment}
\usepackage{enumerate}
\usepackage[title]{appendix}%
\usepackage[font=small,labelfont=bf,justification=justified]{caption}
\usepackage{subcaption}
\usepackage[shortlabels]{enumitem}
\hypersetup{breaklinks=true}
\begin{document}

%\pagestyle{fancy}
%\rhead{\includegraphics[width=2.5cm]{vch-logo.png}}

\title{Emulating the measurement postulates of quantum mechanics via non-Hermitian Hamiltonian}

\author{Gurpahul Singh}
\email{gs18ms106@iiserkol.ac.in}
\author{Ritesh K. Singh}
\email{ritesh.singh@iiserkol.ac.in}
\author{Soumitro Banerjee}
\email{soumitro@iiserkol.ac.in}
\affiliation{Department of Physical Sciences, Indian Institute of Science Education and Research Kolkata, Mohanpur, 741 246, India }%\noaffiliation

\date{\today}% It is always \today, today,
             % but any date may be explicitly specified

\begin{abstract}
%The measurement problem and the collapse of the wavefunction have been greatly debated issues since the foundation of quantum mechanics.
Ever since the formulation of quantum mechanics, there is very little understanding of the process of the collapse of a wavefunction. We have proposed a dynamical model to emulate the measurement postulates of quantum mechanics. We postulate that a non-Hermitian Hamiltonian operates during the process of measurement, which evolves any state to an attracting equilibrium state, thus, mimicking a ``collapse". We demonstrate this using a 2-level system and then extend it to an N-level system. For a 2-level system, we also demonstrate that the dynamics generated by the Lindblad master equation can be replicated as an incoherent sum of the evolution by two separate non-Hermitian Hamiltonians. %This model presents a simplified approach to capture some of the features of the collapse theory.

\end{abstract}

%Use showkeys class option if keyword
                              %display desired
\maketitle

%\tableofcontents

\section{Introduction}
Any standard quantum mechanics textbook \cite{sakurai_napolitano_2017, nielsen_chuang_2010} begins with a set of postulates. Three of them are termed the ``measurement postulates". % which we are currently interested in. 
These are:
\begin{enumerate}[start=1, label={P\arabic*.}]
    \item A wavefunction, upon measurement of an observable, will always ``collapse" to one of the eigenstates of the operator corresponding to that observable.
    \item The probability of collapse to a particular eigenstate depends on the probability amplitudes associated with each eigenstate in the linear decomposition of the wavefunction.
    \item Any repeated measurement immediately after a collapse gives the same eigenstate.
\end{enumerate}
%A wavefunction, upon measurement corresponding to an operator, will always ``collapse" to one of the eigenstates of the operator.
%The third one is valid only in an energy measurement.
In the case of an energy measurement, a closed quantum system continues to be in the collapsed energy eigenstate. %Any proposed model of measurement should emulate these postulates as a consequence. 

For almost a century, physicists and philosophers have tried to come up with some kind of motivation for them. 
A particular aspect of the measurement problem---the collapse of the wavefunction---has been an area of active research. %Numerous models have been proposed to explain its nature. 
One of the earliest models to explain this phenomenon was the De-Broglie Bohm theory \cite{BohmQM}, %the ``hidden variable theory" that stemmed from the EPR paper. 
a non-local hidden variable theory. Other models that followed were Everett many-worlds interpretation \cite{EverettManyworld}, von Neumann-Wigner interpretation \cite{Wigner'sfriend}, Ghirardi-Rimini-Weber (GRW) theory \cite{GRWmodel}, Continuous Spontaneous Localization (CSL) model \cite{CSLmodel1, CSLmodel2}, Diosi-Penrose (DP) model \cite{Diosi, Penrose1996}, Relational QM \cite{RelationalQM} and decoherence model of collapse \cite{Decoherence}. Through these models, however, no light has been thrown on what happens {\em during} collapse. The experimental tests of the Bell's inequality \cite{Bellsinequality, Bellsinequalitytest1, Bell'sinequalitytest2Aspect} and the CHSH inequality \cite{CHSH} have shown that the process of collapse cannot follow a local hidden variable theory.

We propose that the measurement process occurs over a finite interval. During this interval, the apparatus interacts with the system, and the whole dynamics is governed by a Hamiltonian acting on the combined Hilbert space of the system and the apparatus. This leads to \textit{leaking} of probabilities from the system to the apparatus and vice-versa. When seen from the point of view of the system Hilbert space alone, the dynamics appears to be governed by a non-Hermitian Hamiltonian.

Non-Hermitian Hamiltonians with $\mathcal{PT}$-symmetry (where $\mathcal{P}$ is the reflection operator in space and $\mathcal{T}$ is the time-reversal operator) have been of interest ever since their introduction in the late 1990s \cite{FirstpaperonPT, SecondpaperonPT}.
There has been a growing number of experiments associated with these non-Hermitian Hamiltonians \cite{ExperimentsPT, PTreview2}. Physicists have tried to explore the link between weak measurement and non-Hermitian operators \cite{WeakmeasurementnonHemitian3, WeakmeasurementnonHermitian, WeakmeasurementnonHemitian2}. The connection between master equation Lindblad-Kossakowski type \cite{Lindblad1, Lindblad2, Lindblad3} and non-Hermitian or pseudo-Hermitian dynamics has been studied before \cite{LindbladandnonHermitian, LindbladandnonHermitian2, Lindbladconn}.  

\begin{figure*}[t]
    \centering
    \begin{subfigure}[b]{0.3\textwidth}
        \includegraphics[width = \textwidth]{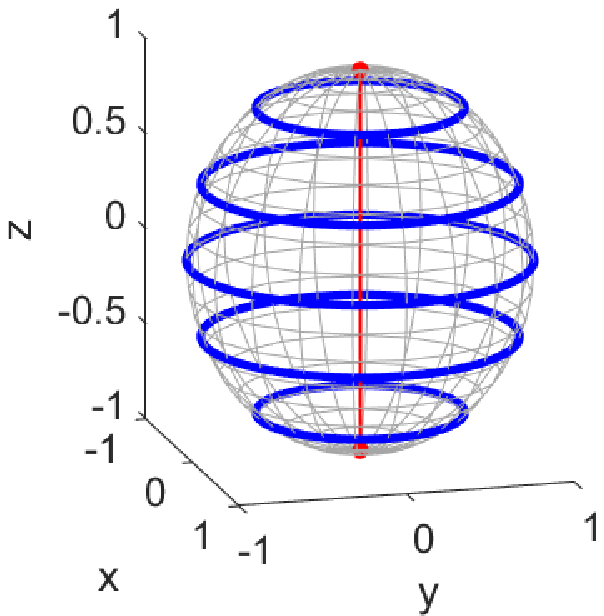} 
        \caption{}
        \label{fig:1a}
    \end{subfigure}
    \begin{subfigure}[b]{0.3\textwidth}
        \includegraphics[width = \textwidth]{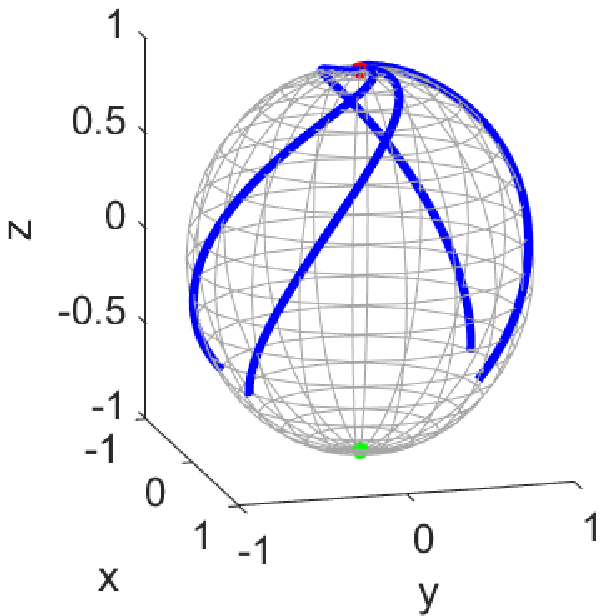}
        \caption{}
        \label{fig:1b}
    \end{subfigure}
    \caption{\small\justifying{{(a) Bloch sphere trajectories (blue) of different initial states evolved via $H = \sigma_z$ using Eq. \eqref{eq:02}. The states (all pure) oscillate on the surface of the Bloch sphere in the $x$-$y$ plane. The red line indicates the fixed points inside the Bloch sphere along the z-axis. (b) For $\gamma = 3$, Bloch sphere trajectories spiral toward the sink at the North pole (red) while the source sits at the South pole (green).}}}
    \label{fig:1}
\end{figure*} 

%If a non-Hermitian Hamiltonian has an antilinear symmetry (such as $\mathcal{PT}$-symmetry), then its eigenvalues are real or appear as complex conjugate pairs \cite{Antilinear2}. 
For our model, we use a non-Hermitian measurement Hamiltonian $H_m$ to evolve a state $\rho$ which satisfies a trace-preserving nonlinear von Neumann equation~\cite{NonlinearvonNeumann}. If $H_m$ has complex eigenvalues, then the state under time evolution would reduce to the eigenvector of $H_m$ having the largest imaginary part of the eigenvalue. In other words, the eigenvector with the largest imaginary part of the eigenvalue will become the attractor of the dynamics. %One can use this property of non-Hermitian Hamiltonians to propose a dynamical model for a state to converge to corresponding eigenvectors, i.e., the wavefunction collapse. This idea has been used to formulate a model of collapse.
This feature helps us emulate the postulates P1 and P3. Such a \textit{deterministic} approach fails to emulate P2 in a simple manner. %For our model to be a valid one, it has to respect the three postulates mentioned previously.  

The organization of the paper is as follows: In Sec.~\ref{II}, the nonlinear von Neumann equation has been introduced that dictates the time evolution of a state when a non-Hermitian Hamiltonian is applied and study its implications. In Sec.~\ref{III}, we lay down the assumptions and the non-Hermitian Hamiltonian that operates in the measurement interval. This Hamiltonian is applied to a two-level system state and it has been shown that one can make any initial state converge to either of the eigenvectors. Then a 4-level state has been considered to study the effect of degeneracy in real and imaginary part of the eigenvalues. Three different diagonal non-Hermitian Hamiltonians have been separately discussed in Sec.~\ref{V}. One of the cases turns out to have a connection to the Markovian dynamics of the Lindblad master equation. In Sec.~\ref{5} and Sec.~\ref{VI}, we discuss the challenges related to our model and summarize the results of our work.

\section{The nonlinear von Neumann equation} \label{II}

Let us consider a non-Hermitian Hamiltonian $H$ of the form $H = H_h - iH_a$ where $H_h$ and $H_a$ are the Hermitian and anti-Hermitian parts. Also, $H_h^{\dagger} = H_h$ and $H_a^{\dagger} = H_a$. The density matrix $\rho$ of a system, evolved via the Hamiltonian $H$ will follow \cite{NonlinearvonNeumann}
\begin{equation}\label{eq:01}
    \dot\rho = -i[H_h,\rho]-\{H_a,\rho\} + 2\:\text{tr}(\rho H_a)\rho
\end{equation}
which is similar to the nonlinear Schr\"odinger equation mentioned in \cite{Gisin}. %The nonlinearity comes in because of the last term on the RHS introduced to preserve the trace of $\rho$. 
The evolved state $\rho(t)$ is then given by
\begin{equation}\label{eq:02}
    \rho(t) = \frac{e^{-iHt}\rho(0)e^{iH^{\dagger}t}}{\text{Tr}(e^{-iHt}\rho(0)e^{iH^{\dagger}t})}
\end{equation}
%where we see a time dependent normalisation factor. 

%For most of our analysis, we will investigate how the state evolves on the Bloch sphere. 
\noindent One can write a general density matrix as 
$$\rho = \frac{1}{2}(\mathbb{I} + x\sigma_x + y\sigma_y+z\sigma_z)$$
where $\sigma_i$ are the Pauli matrices and $x, y, z$ represent the coordinates of the state on (or inside) the Bloch sphere. Each coordinate $i = \text{Tr}(\rho\sigma_i)$ and $x^2 +y^2 +z^2 \leq 1$.

We will be working in the diagonal basis throughout the paper since it does not give rise to any new physics while making the algebra slightly easier.

Let $H = \sigma_z - i\frac{\gamma}{2}(\mathbb{I}-\sigma_z)$ such that $H_h = \sigma_z$ and $H_a = \frac{\gamma}{2}(\mathbb{I}-\sigma_z)$, where $\gamma \geq0$. We put the expression for $\rho, H_h$ and $H_a$ in Eq.~\eqref{eq:01}, multiply with the Pauli matrices and take the trace to obtain:
\begin{subequations}\label{eq:03}
    \begin{align}
        \dot x &= \text{Tr} (\dot \rho\sigma_x) = -2y -\gamma xz \label{eq:03a}\\
        \dot y &= \text{Tr} (\dot \rho\sigma_y) = 2x - \gamma yz \label{eq:03b}\\
        \dot z &= \text{Tr} (\dot \rho\sigma_z) = -\gamma z^2 +\gamma \label{eq:03c}
    \end{align}
\end{subequations}

The bilinear terms in the above equations are coming from the nonlinear term in \eqref{eq:01}. If $\gamma=0$, we see that $\dot z = 0$ which means the Bloch sphere trajectories 
lie in the $x$-$y$ plane. $\dot x = 0, \dot y=0 \Rightarrow x = 0, y = 0$. So, z-axis is a line of fixed points from $z=-1$ to $z=1$. Analysing the stability of the line of fixed points, it is found that all these points are centers. Since the equations are linear in $x$ and $y$, the trajectories are circles everywhere inside and on the Bloch sphere. This can be seen in Fig.~\ref{fig:1a} where only the pure state trajectories have been plotted. $\gamma = 0$ also means that $H$ has real eigenvalues.

\begin{figure*}[t]
    \centering
    \begin{subfigure}[b]{0.28\textwidth}
    \includegraphics[width=\textwidth]{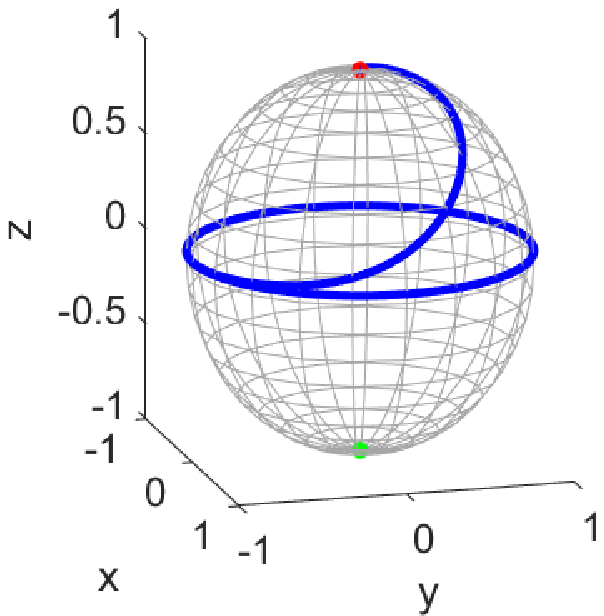}
    \caption{}
    \label{fig:2a}
    \end{subfigure}
    \begin{subfigure}[b]{0.35\textwidth}
    \includegraphics[width=\textwidth]{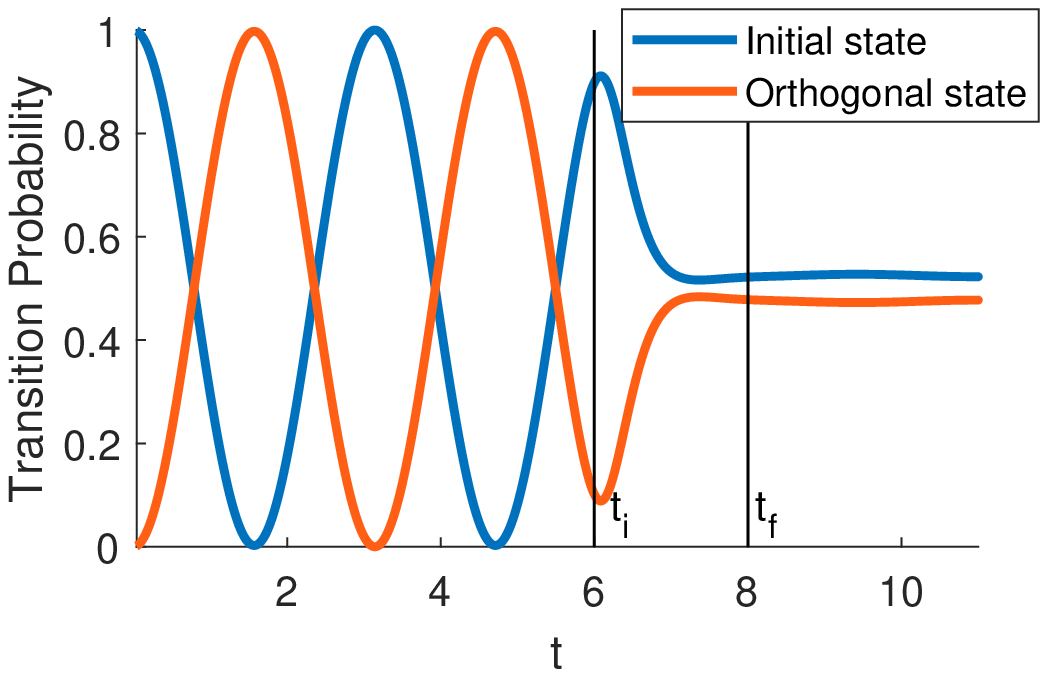} 
    \caption{}
    \label{fig:2b}
    \end{subfigure}
    \begin{subfigure}[b]{0.35\textwidth}
    \includegraphics[width=\textwidth]{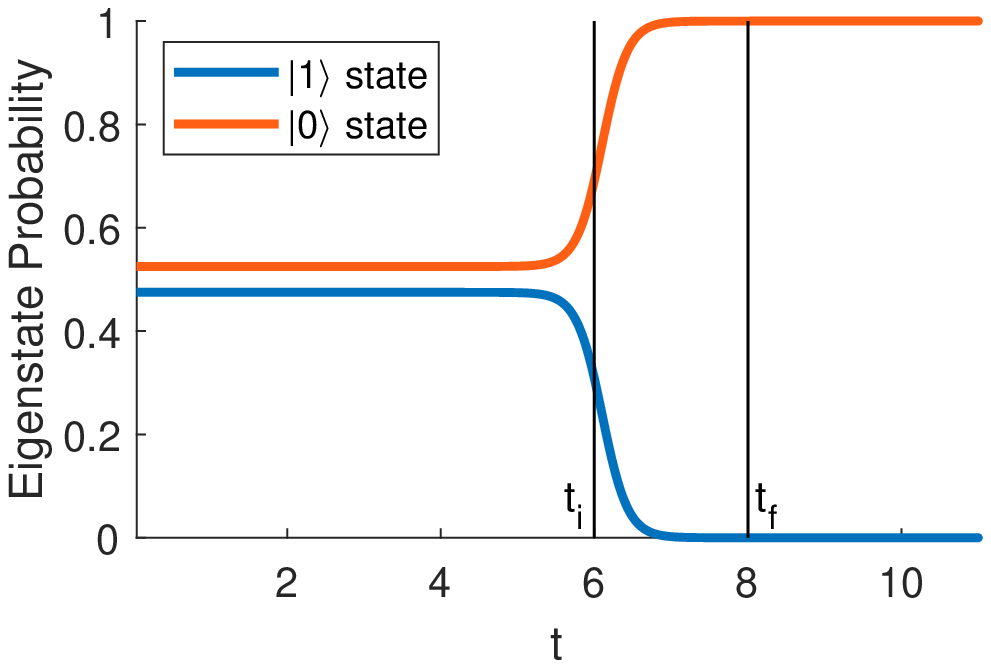}
    \caption{}
    \label{fig:2c}
    \end{subfigure}
    \begin{subfigure}[b]{0.28\textwidth}
    \includegraphics[width=\textwidth]{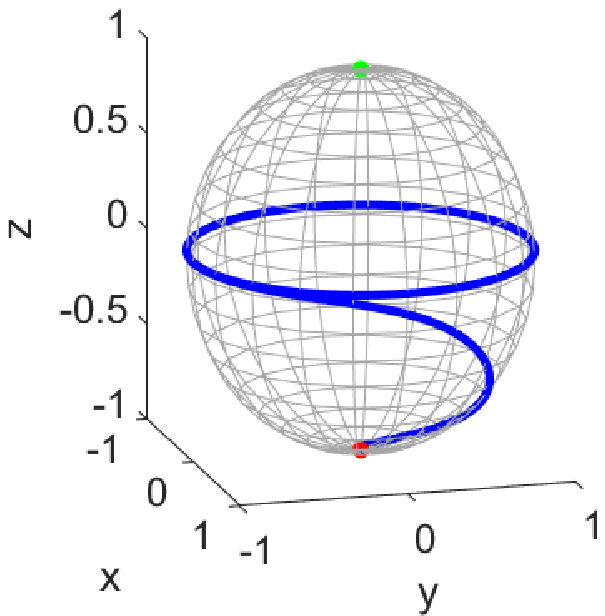} 
    \caption{}
    \label{fig:2d}
    \end{subfigure}
    \begin{subfigure}[b]{0.35\textwidth}
    \includegraphics[width=\textwidth]{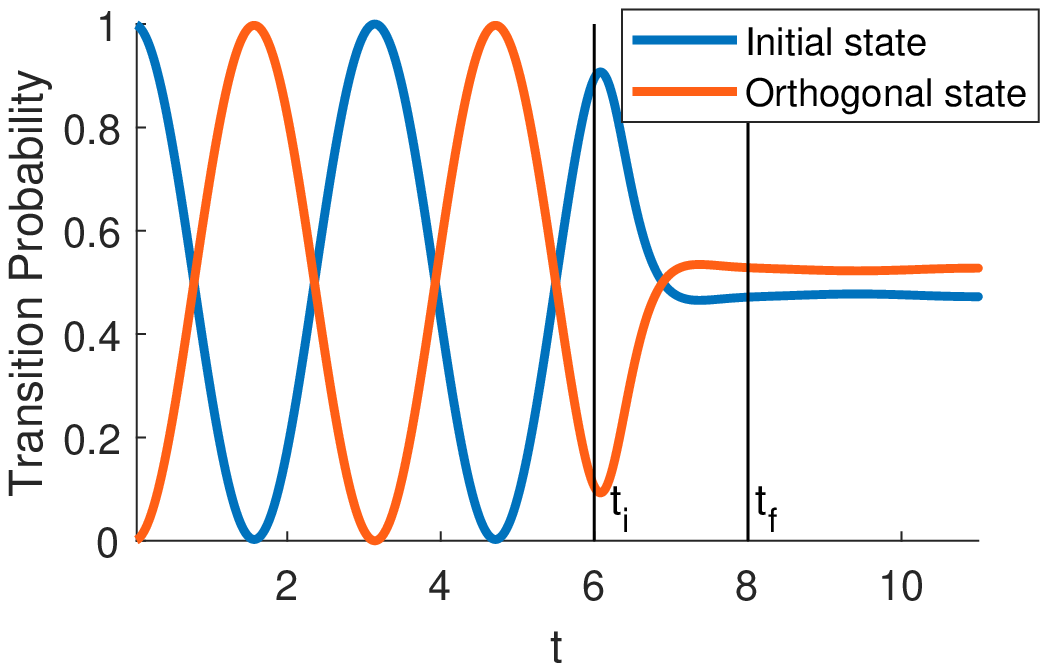} 
    \caption{}
    \label{fig:2e}
    \end{subfigure}
    \begin{subfigure}[b]{0.35\textwidth}
    \includegraphics[width=\textwidth]{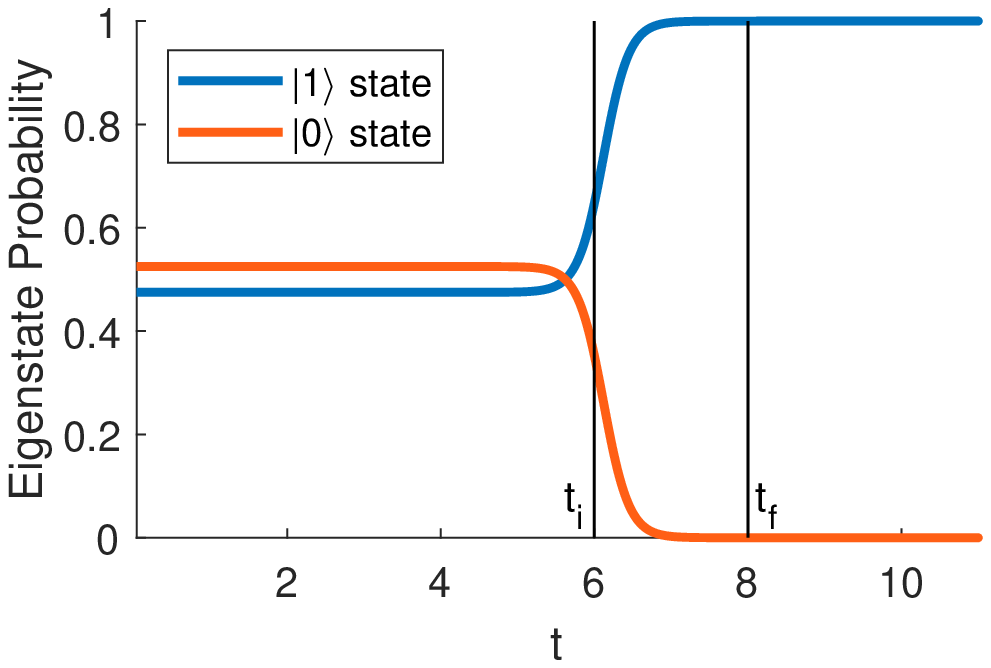}
    \caption{}
    \label{fig:2f}
    \end{subfigure}
    \caption{\small\justifying{{Bloch sphere trajectory of a state with $H=\sigma_x$ before $t_i=7$ and after $t_f = 8$ and with $H_m=i\epsilon\sigma_x$ (a) and $H_m=-i\epsilon\sigma_x$ (d) in between $t_i$ and $t_f$. $\epsilon= \sqrt{\gamma_m^2-1}$ and $\gamma_m = 3$. The red and green markers denote the sink and the source, respectively. [(b) and (e)] Probabilities of being in the initial state (blue) and its orthogonal state (orange) are shown. (c) The probability of being in the $\ket{+}$ state grows to 1 and being in $\ket{-}$ goes to 0, while the opposite happens in graph (f).}}}
    \label{fig:2}
\end{figure*}

The case of interest is of $\gamma \neq 0$. The fixed points come out as 
\begin{eqnarray*}
\dot x, \dot y &=& 0 \Rightarrow x, y = 0 \\
\dot z &=& 0 \Rightarrow z^* = \pm 1
\end{eqnarray*}
which means that the only fixed points in this case are on the Bloch sphere, at the North and the South pole. The stability of these fixed points show that the point $(0,0, -1)$ has positive real eigenvalue in the z-direction and an outward spiralling flow in the $x$-$y$ plane (complex eigenvalues with positive real parts). Thus, this point acts as a source. On the other hand, the point $(0,0,1)$ has negative real eigenvalue in the z-direction and the flow in the $x$-$y$ plane is like an attracting spiral (complex eigenvalues with negative real parts). Thus, this point is a sink. Also, these points correspond to the eigenvectors of $H$ which are $\ket 0$ (the sink) and $\ket 1$ (the source) since $H$ is diagonal. The Bloch sphere trajectories for this case are shown in Fig.~\ref{fig:1b}.

In summary, there are two disjoint parameter regions. For $\gamma=0$, there is a line of fixed points with closed periodic trajectories. For $\gamma>0$, $H$ has complex eigenvalues. Here, one can see a sink and a source existing on the Bloch sphere. %This also means that any initially mixed state would finally end up at the surface of the Bloch sphere (purification). 
The latter case is important for the measurement postulates as will be seen in the next section.

\section{The measurement Hamiltonian}\label{III}

To emulate the measurement postulates, we lay down the following assumptions: 

\begin{enumerate}[start=1, label={A\arabic*.}]
    \item  There is a finite time during which the measurement takes place. We take $t_i$ as the point where this process starts and $t_f$ when it ends.
    %\item There is a parameter which represents the strength of coupling between system and the measuring apparatus.
    \item In between $t_i$ and $t_f$, a ``measurement Hamiltonian" $H_m$ acts on the state. $H_m$ is a non-Hermitian Hamiltonian with complex eigenvalues.  
\end{enumerate}

\begin{figure*}
    \centering
    \begin{subfigure}[b]{0.45\textwidth}
        \includegraphics[width = \textwidth]{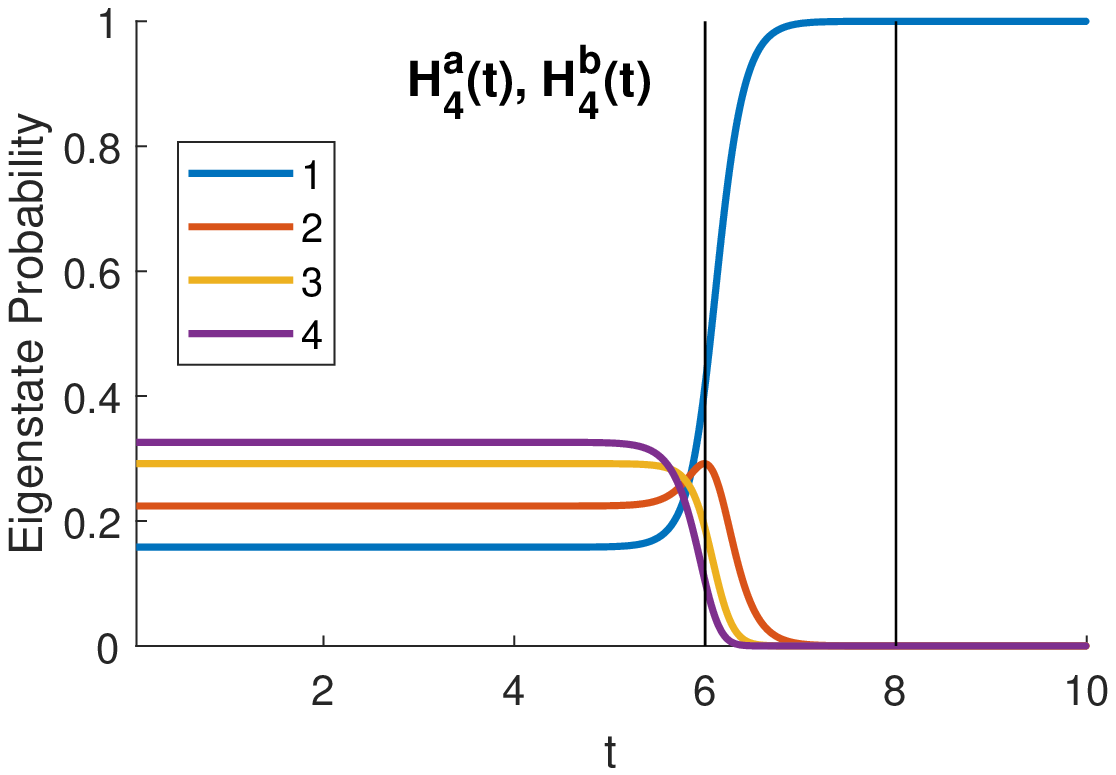} 
        \caption{}
        \label{fig:3a}
    \end{subfigure}
    \begin{subfigure}[b]{0.45\textwidth}
        \includegraphics[width = \textwidth]{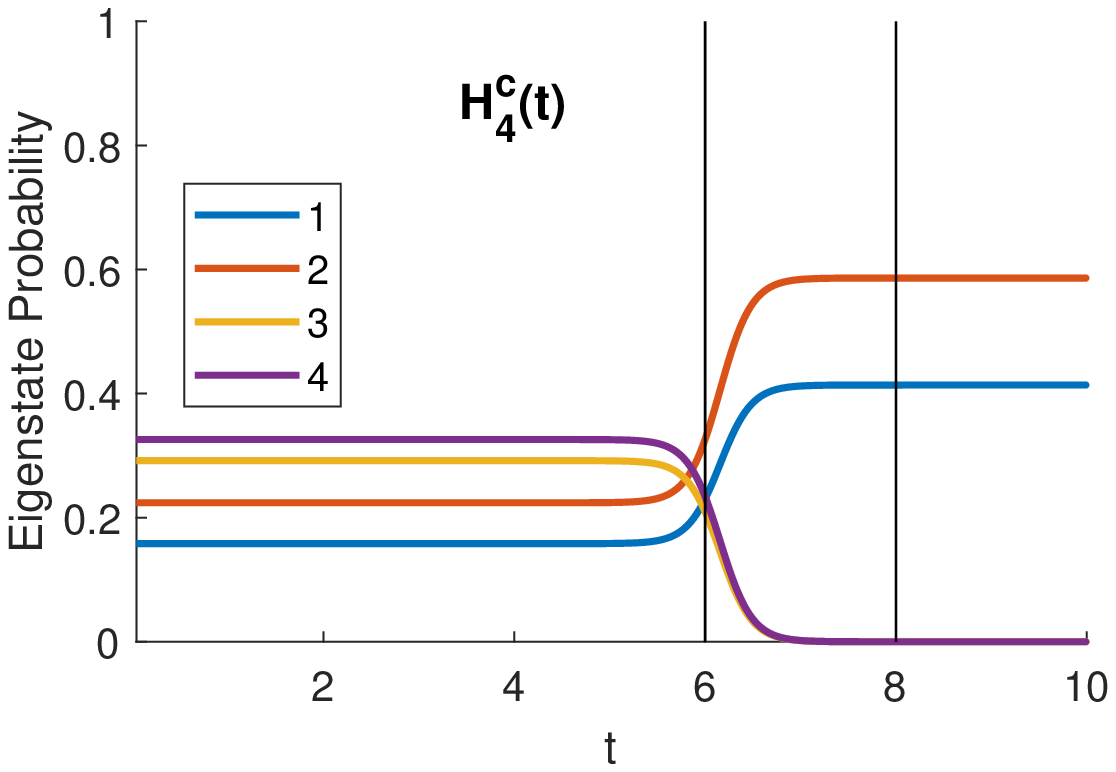}
        \caption{}
        \label{fig:3b}
    \end{subfigure}
    \caption{\small\justifying{{(a) Probability of being in one of the eigenstates for a 4-level state when either (Case a) both real and imaginary parts of the eigenvalues are non-degenerate ($H_4^a(t)$) or (Case b) there is degeneracy only in the real part of the eigenvalues ($H_4^b(t)$). (b) There is degeneracy only in the largest imaginary part of the eigenvalues. $\gamma=3$, $t_i = 6$ and $t_f = 8$.}}}
    \label{fig:3}
\end{figure*}
%. But same is not true for the $PT$ broken case. 

%\begin{figure}
%    \centering
%    \includegraphics[scale=0.64]{sx.eps}
%    \caption{Bloch sphere trajectories (blue) of different initial states evolved via $H = \sigma_x$. The states (all pure) oscillate on the surface of the Bloch sphere in the $x$-$y$ plane. Red dots indicate the eigenvectors of $H$.}
%    \label{fig:1}
%\end{figure}
Before the measurement starts, the state is either a static state (a point on the Bloch sphere) or one that is oscillating on the Bloch sphere due to some Hermitian driving Hamiltonian. Also, any $\mathcal{PT}$-symmetric Hamiltonian with real eigenvalues can be made Hermitian using some similarity transformation \cite{Similarity}. Thus, in general, one can have a $\mathcal{PT}$-symmetric Hamiltonian $H$ with real eigenvalues driving the state before the measurement and then a $\mathcal{PT}$-broken Hamiltonian $H_m$ operates during the measurement from $t_i$ to $t_f$. In that time period, the state would go to one of the eigenvectors of $H_m$ which can be seen from our analysis in section II.

After $t_f$, $H$ acts once again on the state, but because of the measurement process, the state should remain at the eigenvector of $H_m$. For this, the eigenvector to which $H_m$ sends the state must exactly match with the eigenvector of $H$. 

Since eigenvectors of $H$ and $H_m$ match, one can represent both in the diagonal basis without loss of generality. The Hermitian Hamiltonian $H$ in the eigenvector basis can be written as 
\begin{equation}\label{eq:11}
    H = \sum_i \lambda_i \braket{\phi_i}{\phi_i},
\end{equation}
then $H_{m,j}$ required to collapse to $\ket \phi_j$ is given as 
\begin{equation}\label{eq:12}
    H_{m,j}= i\gamma \braket{\phi_j}{\phi_j} + \sum_{i} \lambda_i \braket{ \phi_i}{\phi_i}
\end{equation}

where $\gamma > 0$. Combining Eqs.~\eqref{eq:11} and \eqref{eq:12} into one single time dependent Hamiltonian $H_j(t)$:

\begin{equation}\label{eq:13}
    H_j(t) = \sum_{i} \lambda_i \braket{\phi_i}{\phi_i} + i\gamma f(t)P_j
\end{equation}
where $P_j = \braket{\phi_j}{\phi_j}$ is the projection operator to eigenvector $\ket{\phi_j}$, and $f(t)$ is a switching function which switches on the second term at $t=t_i$ and switches it off at $t=t_f$. We use this function for a smooth transition from Hermitian to non-Hermitian Hamiltonian and back, during a measurement. In the following simulations, we use $f(t)$
\begin{equation}\label{eq:ft}
f(t) = [\tanh(\gamma(t-t_i))-\tanh(\gamma(t-t_f))]/2.    
\end{equation}

We have assumed for simplicity that $\gamma$, which sits in the imaginary part of eigenvalue and decides the speed of collapse, also controls the speed of switching in the switching function.

The degree of collapse to a chosen state can be parameterized as 
$$\kappa = 1-e^{-\gamma(t_f-t_i)}$$ 
which tends to 1 for large enough values of $\gamma(t_f-t_i)$. The speed of convergence is higher for higher $\gamma$ and vice-versa. We can think of $\gamma$ being proportional to system-apparatus coupling strength. That means if the system-apparatus coupling is smaller, the state would collapse slower. Similarly, increasing the measurement interval restores the degree of collapse.

Let us show measurement of spin of a two-level state in the z-direction using our model. 

\subsection{Two-level state}

We consider specific Hamiltonians 
\begin{equation}\label{eq:Ht}
    H_{\pm}(t) =\sigma_z+i\dfrac{\gamma}{2}f(t)(\mathbb{I}\pm\sigma_z)
\end{equation}
$H_+$ and $H_-$ makes an initial state collapse to $\ket 0$ and $\ket 1$ respectively. So, depending on the sign in the Hamiltonian, we have an initial state collapsing to either of the eigenvectors. The graphs are shown in the bottom row in Fig.~\ref{fig:2}.

Once $H_{\pm}(t)$ evolves an initially oscillating state to one of the fixed points, it would stay there even after $t_f$. This is shown in Figs.~\ref{fig:2a} and \ref{fig:2d}. All the plots are generated by taking an initial state and evolving it using \eqref{eq:02}.
In Figs.~\ref{fig:2b} and \ref{fig:2e}, the probability of being in the initial state, i.e., $\text{Tr}(\rho(t)\rho(0))$ is shown in blue and the probability of being in the orthogonal state, $\text{Tr}(\rho(t)(\mathbb{I} - \rho(0)))$, in orange. The probabilities of being in the $\ket 1$ (blue) and in $\ket 0$ state (orange) are shown in Figs.~\ref{fig:2c} and \ref{fig:2f}.
%Thus, given enough time, any state would approach the fixed point asymptotically.

\subsection{Degeneracy}\label{IV}
Let us now see the effect of degeneracy in the real and imaginary parts of the eigenvalues on the collapse using three cases of $N=4$ Hamiltonians. %For dimensions $N>2$, Bloch-sphere-like representations are not possible, so we show only the eigenstate probabilities to demonstrate collapse.  

\textbf{Case a:}  Both the real part and the imaginary part of the eigenvalues are non-degenerate. We choose the Hamiltonian to be
$$H_4^a(t) = \text{dia}(1,2,3,4) + i\gamma f(t)\text{dia}(4,3,2,1)$$
where $\gamma = 3$. Hence, the collapse should happen to the eigenstate with eigenvalue 1 because it has the largest imaginary part as shown in Fig.~\ref{fig:3a}. All other eigenstates, although also having positive imaginary parts, would start decaying to 0 because of the relative difference in the imaginary parts.

\textbf{Case b:} The imaginary parts are non-degenerate while the real parts are degenerate. 
$$H_4^b(t) = \text{dia}(1,1,1,4) + i\gamma f(t)\text{dia}(4,3,2,1)$$
Here, the collapse again happens to the 1st eigenstate, further confirming that convergence is only determined by the largest imaginary part (Fig.~\ref{fig:3a}).

\textbf{Case c:} The real parts are non-degenerate while the imaginary parts are degenerate
$$H_4^c(t) = \text{dia}(1,2,3,4) + i\gamma f(t)\text{dia}(0,0,-1,-1)$$
In this case, there are non-positive imaginary parts. The largest imaginary parts being degenerate means that there is no preferred eigenstate to collapse to. The initial state reduces to a two-dimensional subspace spanned by the first two eigenstates 
as shown in Fig.~\ref{fig:3b}. One can also see that the probabilities for the first two eigenstates get rescaled but the ratio of the probability coefficients remains the same. So, in our model, the collapse would take place only if there is an eigenvalue with a unique largest imaginary part.

\section{Connection to the Lindblad formalism}\label{V}

Let us now see how a time evolved density matrix $\rho(t)$ looks like when evolved using a general non-Hermitian Hamiltonian $H$. For this we use 
$$\rho(0) = \frac{1}{2}(\mathbb{I} + r_x\sigma_x+r_y\sigma_y+r_z\sigma_z) = \frac{1}{2}(\mathbb{I}+\vec{r}\cdot\vec{\sigma})$$
where $r_x,r_y,r_z$ are all real. We choose a Hamiltonian $$H=R_0\mathbb{I}+R_1\sigma_x+R_2\sigma_y+R_3\sigma_z = R_0\mathbb{I} + \vec{R}\cdot\vec{\sigma}$$
where $R_0, R_1, R_2, R_3$ can be complex. Also, 
$$H^{\dagger} = R_0^*\mathbb{I}+\vec{R}^*\cdot\vec{\sigma}.$$ 
%Let $ \sqrt{\vec{R}\cdot\vec{R}} = p$ and $\sqrt{\vec{R}^*\cdot\vec{R}^*} = q$. Making use of the formula $e^{ia(\vec{n}\cdot\vec{\sigma})} = \cos{(a)}\mathbb{I}+\dfrac{i}{\vert \vec{n}\vert }\sin{(a)}(\vec{n}\cdot\vec{\sigma})$ and 
Substituting $H$, $H^{\dagger}$ and $\rho(0)$ in \eqref{eq:02}, one has $\rho(t) = \dfrac{N(t)}{\text{Tr}(N(t))}$ where $N(t)$ is
\begin{align}
\label{eq:14}
    \begin{split}
        N(t) = \frac{1}{2}e^{-i(R_0-R_0^*)t}\{\cos{(pt)}\cos{(qt)}(\mathbb{I}+\vec{r}\cdot\vec{\sigma}) \\+ \frac{i}{q}\cos{(pt)\sin{(qt)}}(\mathbb{I}+\vec{r}\cdot\vec{\sigma})(\vec{R}^*\cdot\vec{\sigma})\\
        -\frac{i}{p}\cos{(qt)\sin{(pt)}}(\vec{R}\cdot\vec{\sigma})(\mathbb{I}+\vec{r}\cdot\vec{\sigma})\\
        +\frac{1}{pq}\sin{(pt)}\sin{(qt)}(\vec{R}\cdot\vec{\sigma})(\mathbb{I}+\vec{r}\cdot\vec{\sigma})(\vec{R}^*\cdot\vec{\sigma})\}
    \end{split}
\end{align}
where $ p = \sqrt{\vec{R}\cdot\vec{R}}$ and $q = \sqrt{\vec{R}^*\cdot\vec{R}^*}$. 
%\begin{align}
%    \label{eq:15}
%    \begin{split}
%        \text{Tr}(N(t)) = e^{-i(R_0-R_0^*)t}\{ \cos{(pt)}\cos{(qt)}\\ +\frac{i}{q}\cos{(pt)\sin{(qt)}}(\vec{r}\cdot\vec{R}^*)
%        -\frac{i}{p}\cos{(qt)\sin{(pt)}}(\vec{R}\cdot\vec{r})\\
%        +\frac{1}{pq}\sin{(pt)}\sin{(qt)}(\vec{R}\cdot\vec{R}^* + i(\vec{R}\cross\vec{r})\cdot\vec{R}^*)\}
%    \end{split}
%\end{align}
Let us take the initial state to be 
$$\rho(0) = \left [
 \begin{array}{cc}
 \vert c_1\vert^2&c_1c_2^* \\
 c_1^*c_2&\vert c_2\vert ^2 \\ 
 \end{array}
 \right ]$$
 so that $r_x = c_1c_2^*+c_1^*c_2, r_y = i(c_1c_2^*-c_1^*c_2), r_z=\vert c_1\vert ^2-\vert c_2\vert ^2$. %And we consider a diagonal $H$ with $R_1 = R_2 = 0 \Rightarrow p=R_3, q=R_3^*$. %In that case, 
%\begin{align}
%\label{eq:16}
%    \begin{split}
%        \text{Tr}(N(t)) &= e^{-i(R_0-R_0^*)t}\{ \cos{(R_3t)}\cos{(R_3^*t)}\\
%        &+\sin{(R_3t)}\sin{(R_3^*t)}\}\\
%        &= e^{-i(R_0-R_0^*)t}\{\cos{((R_3-R_3^*)t)}\}
%    \end{split}
%\end{align}
 
Our results in the previous sections have shown that, broadly, there can be three ways one can send an initial state $\rho$ to one of the eigenvectors of a Hamiltonian $H_m$. 

\textbf{Case A:} We add an imaginary number $i\gamma$ ($\gamma>0$) to one eigenvalue and subtract it from the other. This is like ``pushing" the state towards one eigenvector and at the same time ``pulling" it away from the other. So, $H_m$ would look like
\begin{align}
    \label{eq:17}
    \begin{split}
        H^A_m &= \left [
 \begin{array}{cc}
 \lambda_1+i\gamma&0 \\
 0&\lambda_2-i\gamma \\ 
 \end{array}
 \right ] \\
 &= \lambda_0\mathbb{I} + \left(\omega+ i\gamma\right)\sigma_z ,
    \end{split}
\end{align}
where $\lambda_0 = \frac{\lambda_1+\lambda_2}{2}$ and $\omega = \frac{\lambda_1-\lambda_2}{2}$. So, $R_0 = \lambda_0$ and $R_3 = \omega+i\gamma$. 
%Substituting these in \eqref{eq:16}, we obtain
%\begin{equation}\label{eq:18}
%    \text{Tr}(N(t)) = \cosh{(2\gamma t)}
%\end{equation}
We expand \eqref{eq:14} and divide it by the trace to get
\begin{equation}
 \rho^A(t) \!=\!\! \left [ \!
 \begin{array}{cc}
 \dfrac{\vert c_1 \vert ^2}{ \vert c_1 \vert ^2\!+\! \vert c_2 \vert ^2e^{-4\gamma t}}& \dfrac{c_1c_2^*e^{-i2\omega t}}{ \vert c_1 \vert ^2e^{2\gamma t}\!+\! \vert c_2 \vert ^2e^{-2\gamma t}} \\
 \dfrac{c_1^*c_2e^{i2\omega t}}{ \vert c_1 \vert ^2e^{2\gamma t}\!+\! \vert c_2 \vert ^2e^{-2\gamma t}} & \dfrac{ \vert c_2 \vert ^2}{ \vert c_1 \vert ^2e^{4\gamma t}\!+\! \vert c_2 \vert ^2}. \\ 
 \end{array}
 \!\! \right ]
\end{equation}
It is clear from the above equation that as $t\rightarrow \infty$, the off-diagonal terms go 0 (decoherence). The $\rho^A_{22}(t)$ element also goes to 0 while the $\rho^A_{11}(t)$ term goes to 1. Hence the state ultimately approaches $\ket 0$ state given enough time. %Also, the rate of growth and decay depends on the value of $\gamma$. %If we would have added $i\gamma$ to the lower diagonal element of $H^A_m$, then the opposite would have happened, i.e., the state would have gone to $\vert 1\rangle$.

\textbf{Case B:} We add $i\gamma$ to one of the diagonal elements but do not subtract it from the other diagonal. %In a sense, this is ``pushing" the state towards one eigenvector without pulling it from the other. 
$H^B_m$ looks like
$$H^B_m = \left [
 \begin{array}{cc}
 \lambda_1+i\gamma&0 \\
 0&\lambda_2 \\ 
 \end{array} 
 \right ] =\left(\lambda_0+\frac{i\gamma}{2}\right)\mathbb{I} + \left(\omega+\frac{i\gamma}{2}\right)\sigma_z.$$

So, $R_0 = \lambda_0+\frac{i\gamma}{2}$, $R_3 = \omega+\frac{i\gamma}{2}$. in this case, $\rho^B(t)$ turns out to be
\begin{equation}
    \rho^B(t) = \left [ 
 \begin{array}{cc}
 \dfrac{ \vert c_1 \vert ^2}{ \vert c_1 \vert ^2+ \vert c_2 \vert ^2e^{-2\gamma t}}& \dfrac{c_1c_2^*e^{-i2\omega t}}{ \vert c_1 \vert ^2e^{\gamma t}+ \vert c_2 \vert ^2e^{-\gamma t}}\\
 \dfrac{c_1^*c_2e^{i2\omega t}}{ \vert c_1 \vert ^2e^{\gamma t}+ \vert c_2 \vert ^2e^{-\gamma t}}& \dfrac{ \vert c_2 \vert ^2}{ \vert c_1 \vert ^2e^{2\gamma t}+ \vert c_2 \vert ^2} \\ 
 \end{array}
  \right ]
\end{equation}
which is exactly same as $\rho^A(t)$ except that $\gamma$ is replaced by $\gamma/2$. So, the rate of convergence is halved. 

\textbf{Case 
 C$_1$:} Let us subtract $-i\gamma$ from the lower diagonal element. In this case
$$H^C_m =\left [
 \begin{array}{cc}
 \lambda_1&0 \\
 0&\lambda_2-i\gamma \\ 
 \end{array}
 \right ] = \left(\lambda_0-\frac{i\gamma}{2}\right)\mathbb{I} + \left(\omega+\frac{i\gamma}{2}\right)\sigma_z.$$
For this case, $\rho^{C_1}(t) = \rho^B(t)$ which means there is no difference in just pushing a state to an eigenvector or just pulling it away from the other.

\textbf{Case C$_2$:} In this case, the Hamiltonian used is $H^C_m$ again. If one does not require the trace preservation, Eq.~\eqref{eq:01} would not have the last term. Also, the Eqs.~\eqref{eq:03} would not contain the nonlinear terms and would look like:
\begin{subequations}\label{eq:modeq}
    \begin{align}
        \dot x &= -2y -\gamma x \label{eq:modeqa}\\
        \dot y &= 2x - \gamma y \label{eq:modeqb}\\
        \dot z &= -\gamma z +\gamma w \label{eq:modeqc}\\
        \dot w &= \gamma z -\gamma w \label{eq:modeqd}
    \end{align}
\end{subequations}
where $w$ represents the trace of the density matrix. The fixed point in this case is given by $(0,0,z^*,w^*)$ where $z^*=w^*$. The flow near the fixed point in the $x$-$y$ plane remains the same while in the $z$-$w$ plane, it is either static or an attractor. Since, we are starting from a state where trace of $\rho$ is 1, it is expected that the trace would decrease in time till the trace and the z-coordinate become equal.

The density matrix evolution in this case is $\rho^{C_2}(t) = N(t)$ (see Eq.~\eqref{eq:14}) which looks like 
\begin{equation}
     \rho^{C_2}(t) = \left [ 
 \begin{array}{cc}
  \vert c_1 \vert ^2& e^{-\gamma t}c_1c_2^*e^{-i2\omega t} \\
 e^{-\gamma t}c_1^*c_2e^{i2\omega t}&  \vert c_2 \vert ^2e^{-2\gamma t} \\ 
 \end{array}
  \right ]
\end{equation}

Here, all the terms go to 0 as $t\rightarrow \infty$ except $\rho^{C_2}_{11}(t)$ term which stays at $ \vert c_1 \vert ^2$. Therefore, if the density matrix represented an ensemble of particles, this evolution would leave the population in the ground state intact. There would be decay in the excited state population and also decoherence. %If we would have subtracted $-i\gamma$ from the 1st element of $H_m(t)$, the only term remaining in the density matrix after evolution for a long time would be the excited state population.

The Lindblad master equation formalism \cite{Lindblad1, Lindblad2, Lindblad3} gives the evolution of a system state $\rho_s$ which is in contact with the environment. As a result of this interaction, there is decoherence such that given enough time, the density matrix state thermalizes to a mixed state with no off-diagonal terms. The diagonal terms sum up to one and each of those gives the population at different energy levels.   

The C$_2$ case leaves out a density matrix whose trace is not preserved, which means the particle number is not preserved. The ground state population remains as it is. But if there is a different Hamiltonian with $-i\gamma$ in the upper diagonal, one would be left with just the excited state population in the density matrix. Thus, the sum of these two dynamics has a unit trace and looks like the density matrix obtained from Lindblad-type dynamics.

This means that the Lindblad Markovian dynamics can be obtained as an incoherent sum of two different non-Hermitian Hamiltonian dynamics.

\section{Discussion}\label{5}
The model we have proposed here gives a scheme for emulating postulates P1 and P3, but leaves out P2, i.e., the correlation of the collapse to one of the particular eigenstates with the probability amplitudes in the wavefunction has not been addressed. In other words, this model is a deterministic one. There can be several ways to make the model probabilistic. But all of them lead, one way or the other, to a local hidden variable description \cite{EPRParadox, Bellsinequality}. 

For example, for the two-level case, we can consider that the $\pm$ sign in \eqref{eq:Ht} oscillates between $+$ and $-$ with time at a very high frequency. Let $\{t_i=t_0, t_1, t_2, \dots, t_f\}$ be the partition of measurement time from $t_i$ to $t_f$ such that the ratio of these time intervals depends on the ratio of amplitude coefficients in the wavefunction. As before, $|\psi(0)\rangle = c_1\vert0\rangle + c_2\vert1\rangle$. We choose $\dfrac{t_{2n+1}-t_{2n}}{t_{2n+2}-t_{2n+1}} = \dfrac{\vert c_1\vert^2}{\vert c_2\vert^2}$ where $n=0,1,2,\dots$ $\frac{N}{2}-2$. So, the full measurement Hamiltonian would look like
\begin{equation}
    H(t) =\sigma_z+i\dfrac{\gamma}{2}f(t)(\mathbb{I}+g(t)\sigma_z)
\end{equation}
where $f(t)$ is same as before \eqref{eq:ft} and 
$$g(t) = \vert c_1\vert^2 - \vert c_2\vert^2+\sum_{m=1}^{\infty}\left[a_m \cos({\frac{m\pi t}{L}})+b_m \sin({\frac{m\pi t}{L}})\right].$$ 
The Fourier expansion coefficients are
$$a_m = \frac{2}{m\pi}\left[\sin{\left(\frac{m\pi}{L}t_1\right)}-(-1)^m\sin{\left(\frac{m\pi}{2L}(t_0+t_2)\right)}\right],$$ 
$$b_m = \frac{2}{m\pi}[-\cos{\left(\frac{m\pi}{L}t_1\right)}+(-1)^m\cos{\left(\frac{m\pi}{2L}(t_0+t_2)\right)}]$$
and $L = \frac{t_f-t_i}{N}$. It can be clearly seen that the measurement Hamiltonian $H(t)$ is a nonlinear function of the initial state $\psi(0)$, which is included in the expression of $g(t)$. The job of $g(t)$ is straight-forward: It is +1 for the time intervals $t_{2n}$ to $t_{2n+1}$ and $-1$ for time intervals $t_{2n+1}$ to $t_{2n+2}$. That is, in the former interval, the Hamiltonian pushes the state to $\vert0\rangle$ while in the latter interval, it pushes it towards $\vert1\rangle$. 

For a successful measurement, the coupling between the system and apparatus, $\gamma$, must be appropriately large. In fact, $\gamma(t_n-t_{n-1})>1$, and since the intervals are smaller (depending on the frequency), $\gamma$ must be very large. The state oscillates from $\vert0\rangle$ to $\vert1\rangle$ till $t_f$, which is when the oscillation ends, and the state stays at the eigenvector where it was last. 

Thus, in this setup, the final eigenstate for a single particle depends on two factors --- the time when the measurement ends, $t_f$, and the frequency of the oscillation of $g(t)$. One (or both) of these is definitely indeterminable since otherwise, QM would not be probabilistic. And so, we can say that these are local ``hidden variables". The frequency of $g(t)$ might be specific to each prepared state and would result in a different outcome as a result which is ensured by our construction. In other words, $N$ becomes the local hidden variable for this setup that must vary stochastically for each initially prepared state.

We could have chosen some other form for $g(t)$ or maybe even a time-independent one, but would have eventually come to a similar conclusion. Our model becomes probabilistic by only introducing a stochastic local hidden variable into the picture.

\section{Conclusion}\label{VI}
We have given a dynamical model for collapse to a particular eigenstate of the Hamiltonian. We first showed that the evolution of a density matrix via a non-Hermitian Hamiltonian is dictated by a nonlinear von Neumann equation. Its analysis showed that a non-Hermitian Hamiltonian with complex eigenvalues will have an attractor eigenstate---one which has the largest imaginary part of the eigenvalue. Next, we designed a time-dependent diagonal measurement Hamiltonian with a switching function that has the chosen eigenstate as the stable fixed point. %Then, we combined the two Hamiltonians---one operating outside the measurement interval and one inside---into one time-dependent Hamiltonian. %Being written in the eigenvector basis, this Hamiltonian is also diagonal. 
We also showed that the largest imaginary part of the eigenvalues must be unique for collapse to happen. Then we considered three cases of non-Hermitian Hamiltonians and calculated the expressions of time-evolved density matrices. In one of the cases, it was seen that the Lindblad-type evolution in open quantum systems can be obtained as an incoherent sum of two different non-Hermitian Hamiltonian dynamics. Finally, we noted that our measurement model is indeed deterministic and would require a stochastic local hidden variable for it to be probabilistic in nature.

Our model shows what happens in the system subspace when a collapse happens. One can use a Naimark dilation protocol \cite{Dilationbook, Sciencepaper,NaimarkDilation1} to dilate our non-Hermitian Hamiltonian to a higher dimensional Hermitian Hamiltonian that governs the system-ancilla state. Applying this protocol in our model will give us an interaction Hamiltonian that accounts for unitary evolution of the system-ancilla state while the system subspace undergoes non-Hermitian dynamics. This is under investigation and will be reported elsewhere.

\bigskip
\begin{center}
    \textbf{ACKNOWLEDGEMENT}
\end{center}
We would like to thank Sourin Das, Rangeet Bhattacharyya, Anant Varma, and Arnab Acharya for useful discussions. S.B. acknowledges the J.C. Bose National Fellowship provided by SERB, Government of India, Grant No. JBR/2020/000049. G.S. thanks the Department of Science and Technology (DST), Government of India, for support through an INSPIRE Fellowship.

\medskip

%\Urlmuskip=0mu plus 1mu\relax
%\bibliographystyle{}
%\bibliography{main}{}% Produces the bibliography via BibTeX.

\Urlmuskip=0mu plus 1mu\relax
\bibliography{main}{}% Produces the bibliography via BibTeX.

\end{document}